# The result of the Neutrino-4 experiment, sterile neutrinos and dark matter, the fourth neutrino and the Hubble constant

A.P. Serebrov, R.M. Samoilov, M.E. Chaikovskii, O.M. Zherebtsov

NRC "Kurchatov institute" – Petersburg Nuclear Physics Institute, 188300, Gatchina, Russia

serebrov_ap@pnpi.nrcki.ru

**Abstract**

A brief analysis of the result of the Neutrino-4 experiment and the results of other experiments on the search for a sterile neutrino is presented. It is noted that a joint analysis of the results of the Neutrino-4 experiment and the data of the GALLEX, SAGE and BEST experiments confirms the parameters of neutrino oscillations declared by the Neutrino-4 experiment ($\Delta m_{14}^2 = 7.3 \text{ eV}^2$ and $\sin^2 2\theta_{14} \approx 0.36$) and increases the confidence level to $5.8\sigma$.

An estimate of the contribution of sterile neutrinos with these parameters to the energy density of the Universe is made. It amounted to 5.3%. It is discussed that the extension of the neutrino model by introducing two more heavy sterile neutrinos in accordance with the number of types of active neutrinos will make it possible to explain the large-scale structure of the Universe and bring the contribution of sterile neutrinos to the dark matter of the Universe to the level of 27%. Possible contradictions between the measured sterile neutrino parameters and cosmological constraints are analyzed. It is shown that, based on modern astrophysical data, it is impossible to draw a definite conclusion in favor of the model of three or four neutrinos. It is noted that the introduction of the fourth neutrino removes the Hubble tension problem without contradictions with cosmic microwave background observed by the Planck collaboration. The influence of lepton asymmetry on the comparison of models of three or four neutrinos is considered. An estimate was made for the upper limit of the lepton asymmetry $L_e < 0.02$. The possibility of the appearance of lepton asymmetry due to CP violation during oscillations into sterile neutrinos is discussed.

## 1. Introduction

There are quite a few indications of the possibility of the existence of a sterile neutrino. Anomalies were observed in several accelerator and reactor experiments: LSND at a confidence level of 3.8 σ [1], MiniBooNE 4.7 σ [2], reactor anomaly (RAA) 3σ [3,4], as well as in experiments with radioactive sources GALLEX/GNO, SAGE (gallium anomaly - GA 3.2σ) and BEST [5-7]. A detailed comparison of the results of the Neutrino-4 experiment [8] with the results of other experiments is presented in our paper [9]. Here we analyze the result of the Neutrino-4 experiment in connection with the possible role of sterile neutrinos in cosmology. In our previous work [10], the question of cosmological restrictions on sterile neutrinos was raised. In this work, we try to answer the previously posed questions. But we should start by presenting the result of the Neutrino-4 experiment (Fig.1) and jointly analyzing the results of the Neutrino-4 experiment and the data from the GALLEX, SAGE, and BEST experiments.

Comparison of the results of the Neutrino-4 experiment and the result of the BEST experiment and GA is shown in Fig. 2 at the top. An analysis was also performed based on the data of the GALLEX, SAGE, and BEST experiments published in [9]. Using this result together with the result of the Neutrino-4 experiment, the obtained distribution $\Delta\chi^2(\Delta m_{14}^2, \sin^2 2\theta_{14})$ shown in Fig. 2 at the bottom. The parameter value at the best fit point is $\sin^2 2\theta_{14} = 0.38$, $\Delta m_{14}^2 = 7.3 \text{эB}^2$. The confidence level of the observation of oscillations obtained as a result of the joint analysis of the data was 5.8σ.

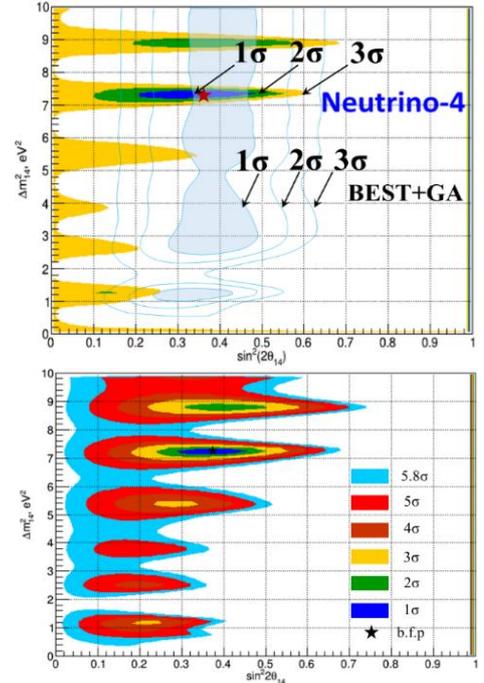

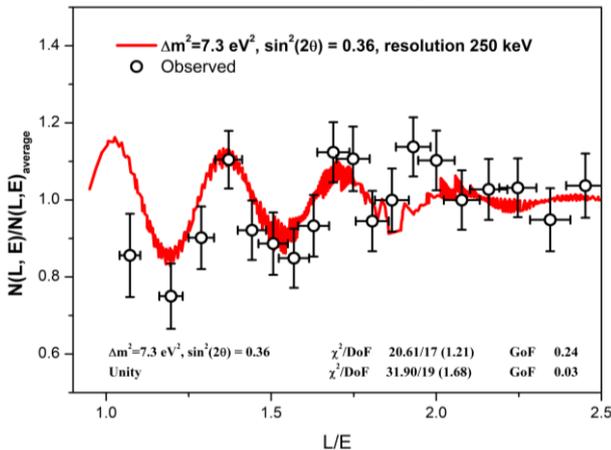

Fig. 1. Oscillatory curve of the neutrino signal. The red line is the expected dependence at $\Delta m_{14}^2 = 7.3 \text{eV}^2$, $\sin^2 2\theta = 0.36$, energy resolution is 250 keV. $\chi^2/DoF = 20.61/17 (1.21)$, $\text{GoF} = 0.24$ for $m_{14}^2 = 7.3 \text{eV}^2$, $\sin^2 2\theta = 0.36$, $\chi^2/DoF = 31.90/19 (1.68)$, $\text{GoF} = 0.03$ for the no oscillation hypothesis.

Fig. 2. On the top - comparison of the result of the BEST experiment together with GA and the result of the Neutrino-4 experiment. At the bottom - the result of a joint analysis of GA, BEST and Neutrino-4, where blue indicates the area with a confidence of 1σ, green - 2σ, yellow - 3σ, dark red - 4σ, red - 5σ and light blue - 5.8σ.

## 2. The role of sterile neutrinos in cosmology

The process of neutrino oscillations in matter changes due to the interaction of neutrinos with matter. This process is especially manifested in primordial plasma.

Interactions of neutrinos with primordial plasma significantly suppress the oscillation process, especially in the early stages. The effective mixing matrix gradually changes from a diagonal matrix at $t = 10^{-5}$ s to a form almost identical to the vacuum mixing matrix at $t = 1$ s.

Now we are interested in the densities of sterile, tau, muon and electron neutrinos at different times. To evaluate them, it is necessary to consider the dynamics of the processes of creation and annihilation of various types of neutrinos by solving a differential equation. We consider the behavior of neutrinos in the era after the annihilation of baryons and antibaryons. During this epoch, the processes of scattering by electrons, positrons, neutrinos, and antineutrinos, as well as the process of annihilation of neutrinos and antineutrinos, contribute to neutrino interactions.

Three processes influence the density dynamics of a sterile neutrino: 1) the expansion of the universe, 2) transitions of an active neutrino into a sterile one, and 3) reverse transitions of a sterile neutrino into an active state. The reverse transition of a sterile neutrino is considered as an oscillation of the sterile state into an active one, followed by the interaction of the active component.

Below is an equation that takes into account the generation of sterile neutrinos $\nu_s$ and their "sink". Equation (1) includes the effective interaction of a sterile neutrino with plasma due to oscillations:

$$\frac{dn_{\nu_s}}{dt} + 3Hn_{\nu_s} = \frac{1}{2}\left(\frac{\sin^2 2\theta_{m\,14}\,n_{\nu_e}}{\tau_{\nu_e}} + \frac{\sin^2 2\theta_{m\,24}\,n_{\nu_\mu}}{\tau_{\nu_\mu}} + \frac{\sin^2 2\theta_{m\,34}\,n_{\nu_\tau}}{\tau_{\nu_\tau}}\right) - \frac{1}{2}\left(\frac{\sin^2 2\theta_{m\,14}}{\tau_{\nu_e}} + \frac{\sin^2 2\theta_{m\,24}}{\tau_{\nu_\mu}} + \frac{\sin^2 2\theta_{m\,34}}{\tau_{\nu_\tau}}\right)n_{\nu_s} \quad (1)$$

where $H$ – Hubble parameter, $1/\tau_{\nu_\alpha}$ – collision frequency for neutrinos of flavor $\nu_\alpha$, $n_{\nu_s}$, $n_{\nu_e}$, $n_{\nu_\mu}$ и $n_{\nu_\tau}$ are the densities of sterile, electron, muon, and tau neutrinos corresponding to the Fermi-Dirac distribution with zero chemical potential [11]. We used the following values for the squared sines of the double angle $\sin^2 2\theta_{14} = 0.36$, $\sin^2 2\theta_{24} = 0.024$ и $\sin^2 2\theta_{34} = 0.043$ from [10].

Equation (1) is an approximation that does not take into account the effect of transitions of active neutrinos into sterile ones on the density of active neutrinos themselves and considers only the transition to sterile neutrinos followed by the transition of sterile neutrinos to active ones. As the initial conditions, it is chosen that the density of sterile neutrinos at the initial moment of time is equal to zero.

This equation can be applied up to the neutrino freeze-out temperature, that is, up to the temperature at which the density decreases so much that the interaction of the neutrino with the plasma matter can be neglected. At this moment, the interaction of the neutrino with matter stops and only the expansion of the Universe affects the further dynamics of the density.

The consequence of the ongoing processes is a very important result: by the time of all neutrinos freeze-out, the density of sterile neutrinos turns out to be approximately the same as the density of electron neutrinos, just as the densities of tau and muon neutrinos turn out to be the same. This situation is reflected in Fig. 3, which shows the dynamics of densities of neutrinos of various types ratios.

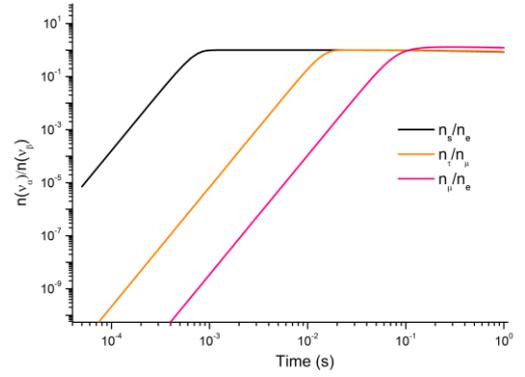

Fig. 3 Neutrino densities ratios.

Now we should estimate the contribution of sterile neutrinos to the energy density of the Universe. It is obvious that the considered sterile neutrino make the main contribution, the density of sterile neutrino particles is the same as the density of active neutrinos, and the mass is much larger- $m_{\nu_4} = 2.7\text{eV}$. The contribution of active neutrinos to the energy density of the Universe is determined by the relation [11]: $\Omega_{\nu_1\nu_2\nu_3} \approx (m_{\nu_1\nu_2\nu_3}/1\text{эВ}) \cdot 0.01h^{-2}$, где $h$ — постоянная Хаббла.

The contribution of the sterile neutrino to the energy density of the Universe is given by:

$$\Omega_{\nu_4} \approx (\sum m_{\nu_i}/1\text{эВ}) 0.01 h^{-2} \cdot n_{\nu_4} m_{\nu_4}/\sum(n_{\nu_i}m_{\nu_i})$$
$$n_{\nu_i} = n_{\nu_e}, \quad \sum(n_{\nu_i}m_{\nu_i}) = n_{\nu_e}\sum m_{\nu_i} \quad (23)$$
$$\Omega_{\nu_4} \approx (2.7\text{eV}/1\text{eV}) \cdot 0.01h^{-2} \cdot 5.1 = 0.053$$

and is 5.3% of the energy density in the Universe.

The same calculations can be carried out for neutrinos with other parameters. Figure 4 shows the dependence of $\Omega_{\nu_4}$ on the sterile neutrino mixing parameters, calculated on the basis of these expressions.

Before we were interested in estimating the contribution of a sterile neutrino to dark matter with parameters close to those obtained in the framework of the Neutrino-4 experiment [8]. However, the above analysis can be extended to the region of higher values of the sterile neutrino mass.

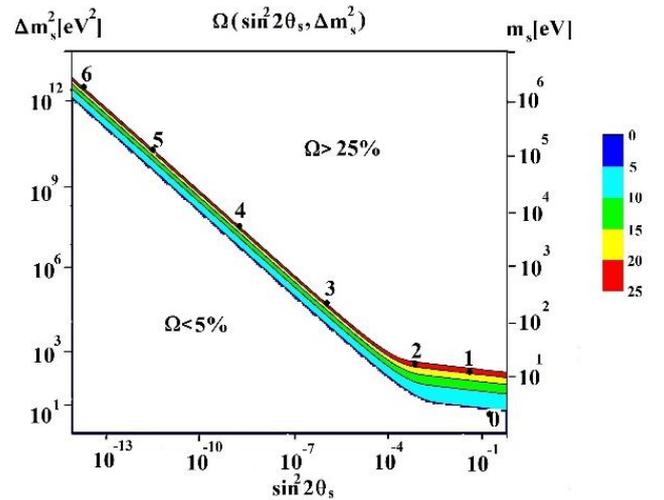

Fig.4 Region of $\Delta m_{14}^2$ and $\sin^2 2\theta_{14}$ leading to allowed contributions to dark matter.

We are interested in the question of what mixing angles for heavy neutrinos leave the contribution of the sterile neutrino to dark matter below the 25% limit. The calculation result is shown in Figure 5. This result shows that heavy sterile neutrinos must have small mixing angles in order not to contradict the cosmological restrictions on the total contribution of dark matter to the energy density in the Universe.

A decrease in the mixing angle leads to the fact that the sterile neutrino does not have time to come into equilibrium with the electron neutrino before the moment of separation of the neutrino from the plasma, that is, the ratio $n_{\nu_s}/n_{\nu_e}$ remains much less than unity. Figure 4 shows 7 points on the plane ($\sin^2 2\theta_{14}, \Delta m_{14}^2$), for which curves of the ratio of the number of sterile neutrinos to the number of electron neutrinos are plotted in Figure 5. As the mixing angle decreases and the mass increases, this ratio decreases.

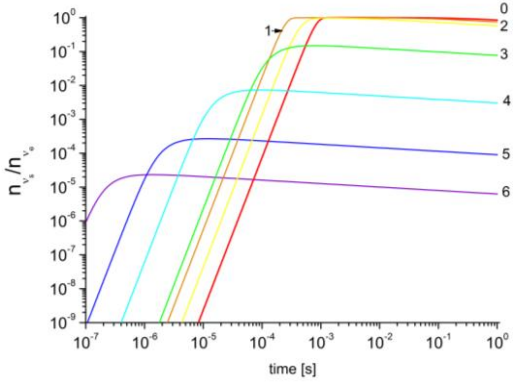

Fig.5. The ratio of the number of sterile neutrinos to the number of electron neutrinos for several values of the parameters on the plane ($\sin^2 2\theta_{14}, \Delta m_{14}^2$). The red line refers to point "0" in figure 4 and corresponds to $\Delta m^2 = 7.3 \text{ эВ}^2, \sin^2 2\theta = 0.36$

From the above analysis, we can conclude that heavy sterile neutrinos should have a small mixing angle due to cosmological constraints. This means that heavy sterile neutrinos do not contribute to reactor neutrino experiments, since at such small mixing angles, they turn out to be unobservable.

The following conclusions can also be drawn from this analysis.

1. A sterile neutrino with parameters $\Delta m_{14}^2 = 7.3 \text{ эВ}^2$, $\sin^2 2\theta_{14} = 0.36$ contributes approximately 5% to dark matter but is relativistic and does not explain the large-scale structure of the Universe.

2. To explain the large-scale structure of the Universe, heavy sterile neutrinos with very small mixing angles are needed.

3. Extension of the neutrino model by introducing two more heavy sterile neutrinos according with the number of types of active neutrinos will make it possible to explain the structure of the Universe and bring the contribution of sterile neutrinos to the dark matter of the Universe to the level of 27%.

Above, we discussed the contribution of the sterile neutrino to dark matter and considered the constraints associated with the total energy of dark matter. We concluded that the parameters obtained in the experiment do not contradict the restriction on the energy density, and moreover, there is room for the introduction of more severe sterile states.

Below we will try to analyze what constraints may arise on the description of dark matter in the Universe using sterile neutrinos.

### 3. Analysis of constraints on sterile neutrinos from laboratory experiments and from cosmology.

There are limitations based on cosmological models and observations. There are three types of observations that can impose restrictions on sterile neutrinos: 1) primordial nucleosynthesis and distribution of light nuclei [12, 13] 2) cosmic microwave background (CMB) [14] 3) clustering of large-scale cosmological structures [15, 16].

The addition of a sterile neutrino with a mass ~1 eV to the model of the evolution of the early Universe changes the number of relativistic degrees of freedom during the period of nucleosynthesis and affects the nature of the expansion of the Universe at the moment of neutrino decoupling. As a result, a sterile neutrino shifts the moment of neutron freeze-out, which means that it affects the ratio of light nuclei in the universe. The anisotropy of the microwave background also turns out to be sensitive to the sterile neutrino parameters. The effect of neutrinos is usually expressed in terms of the effective number of relativistic degrees of freedom $N_{\text{eff}}$. The model with three active neutrinos predicts $N_{\text{eff}}^{3\nu} = 3.046$. The current limit on the effective number of degrees of freedom is derived from fitting data on light elements in the universe $N_{\text{eff}} = 2.843 \pm 0.154$ [17], and microwave background observations lead to $N_{\text{eff}} = 2.99 \pm 0.17$ [14]. It is believed that these results are in good agreement with the 3 active neutrino model and leave open only a limited range of parameters for sterile neutrinos. However, it will be shown later that the restrictions on sterile neutrinos in cosmology are not strict and, moreover, there is still enough room to try to fit sterile neutrinos into the cosmological model.

### 4.1 Experimental constraints on sterile neutrinos from laboratory experiments

Figure 6 shows the same range of possible parameters $\Delta m_{14}^2$ and $\sin^2 2\theta_{14}$, leading to allowed values of the contribution of sterile neutrinos to dark matter.

For sterile neutrinos with energies about several keV, there is a method of laboratory research. The existence of such a sterile neutrino distorts the β-decay spectrum, and therefore can found in experiments on the direct measurement of the electron neutrino mass, based on a detailed study of the β-spectrum in the decay of tritium. At the moment, the best result in experiments of this type was obtained by the KATRIN collaboration [18]. The possibility of establishing experimental limits on eV and keV sterile neutrinos is considered in the KATRIN experiment [19]. Figure 6 shows already excluded (shaded) regions for eV and keV sterile neutrinos, as well as the regions of the limiting sensitivity of the KATRIN experiment. In the range of eV mass sterile neutrinos, KATRIN has the prospect of confirming or refuting our result, however, in the range of keV sterile neutrinos, the sensitivity of the KATRIN experiment is not enough to reach the region where keV sterile neutrinos could be considered as candidates for dark matter particles.

Distortions in the β-decay spectrum are also introduced by sterile neutrinos with a mass of ~ 1eV. In the eV-scale, the limitations obtained in the KATRIN experiment so far [19] (Fig. 6) do not contradict the results of the Neutrino-4 experiment. A further increase in the accuracy in the KATRIN experiment will make it possible to verify the result of the Neutrino-4 experiment. The measurement method used in the KATRIN experiment has a maximum sensitivity in the region of 100–1000 eV$^2$, and in the eV-scale, reactor experiments turn out to be more efficient. A brief analysis of the result of the Neutrino-4 experiment in comparison with other experiments is presented at the beginning of the article, and a detailed analysis can be found in our work [9].

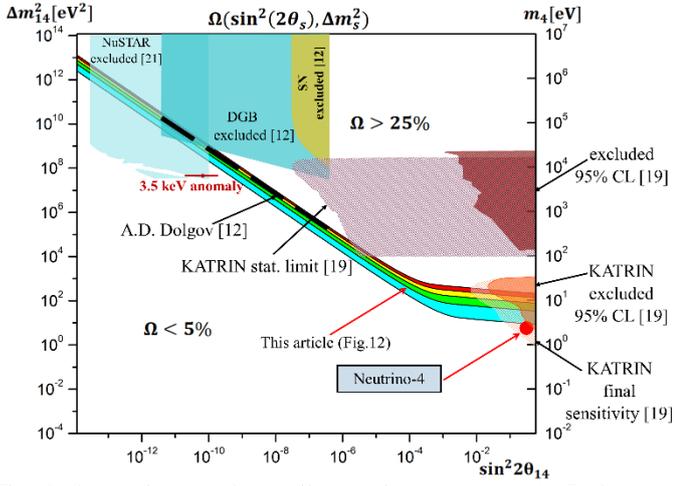

Fig. 6. Constraints on the sterile neutrino parameters.1) Red spot – result of the Neutrino-4 experiment; 2) This article (Fig.4) – area of the $\Omega_s$ values in 5-25% range; 3) A.D. Dolgov [12] – result from the Ref. [12] for $\nu_e - \nu_s$ mixing with $\Omega_s = 30\%$ (it should be noted that result of calculation presented in this work based on equation (1) is consistent with A.D. Dolgov results presented in Ref. [12]); 4) DGB is experimental constraints based on diffuse gamma background; 5) experimental constraints from SN1987 observation; 6) constraints from NuSTAR experiment [21]; 7) KATRIN excluded 95% CL constraints on eV-scale sterile neutrino from KATRIN experiment; 8) KATRIN final sensitivity – sensitivity limit of the KATRIN experiment for eV-scale sterile neutrino; 9) excluded 95% CL – constraints from neutrino mass measurements experiment from Ref. [19]; 10) KATRIN stat. limit[19] – statistical limit of the KATRIN experiment for keV-scale sterile neutrino

In models with a heavy unstable neutrino that decays as a result of mixing with active neutrinos, restrictions on the masses and mixing angles for heavy neutrinos appear. The decay of a heavy neutrino into a light neutrino and a gamma quantum creates radiation of a certain energy equal to $m_s/2$. The mixing angle in this case determines the intensity of the decay, and the concentration of such neutrinos at the moment of neutrino freeze-out, and therefore, ultimately determines the intensity of the emitted gamma quanta. Observations of the gamma-ray spectrum in the range of several tens of keV constraint decaying neutrinos [20]. These limitations are also shown in Fig. 6. The result of experimental observations, the so-called 3.5 keV anomaly, which was closed by subsequent experiments [21], is also shown here.

**4.2 Constraints on the sum of masses of active neutrinos from cosmology.**

Now consider the limit on the sum of masses of active neutrinos from cosmology. It must be emphasized right away that this is a limitation on the sum of active neutrino masses. The question is what this has to do with sterile neutrinos:

$$N_{\text{eff}} = 2.96^{+0.17}_{-0.16}, \sum m_\nu < 0.12 \text{ eV}$$

The point is that this constraint from [22] is related to the requirement for the formation of large-scale structures. In our scheme with three sterile neutrinos, two heavy neutrinos are responsible for the formation of dark matter, and they do not thermalize. A light sterile neutrino with a mass of 2.7 eV is thermalized and participates in the nucleosynthesis process, but by the recombination epoch (T = 0.26 eV), it becomes nonrelativistic and does not contribute to the number of degrees of freedom. Therefore, the number of degrees of freedom at the time of recombination is (as well as without a sterile neutrino):

$$g_*^{rec} = 3.36$$

and remains the same for all following stages of the evolution of the Universe. Thus, the fourth light neutrino mainly affects the process of neutron freeze-out and nucleosynthesis, and then is included in the process of formation of dark matter, where its contribution is 15–20%.

Apparently, a concrete answer to this question requires direct simulation of a model with two heavy and one light neutrino.

**4.3 Experimental limits on sterile neutrinos from astrophysical data on measuring the mass content of $^4$He and other light elements**

Astrophysical measurements can be divided into the following components. Measurement of the mass fraction of $^4$He and other light elements to study the process of primordial nucleosynthesis, measurement of the speed of the galaxies motion to determine the Hubble constant and the study of microwave background, which provides information about the early stage of the evolution of the Universe.

Let's start with the process of nucleosynthesis and turn to the calculation of the $^4$He mass fraction using the well-known calculation scheme from [11].

$$^{4\text{He}}Y = m_{^4\text{He}} \cdot \frac{n_{^4\text{He}}(T_{ns})}{m_p\left(n_p(T_{ns}) + n_n(T_{ns})\right)} = \frac{2}{n_p(T_{ns})/n_n(T_{ns}) + 1}$$

$$n_n(T_{ns})/n_p(T_{ns}) = n_n/n_p \cdot e^{-t_{ns}/\tau_n} = e^{-(m_n-m_p)/T_n} \cdot e^{-t_{ns}/\tau_n}$$

The mass fraction of $^4$He is determined by the ratio of neutrons to protons at the moment of neutron freeze-out at a cosmological time of approximately 1.2 s and a nucleosynthesis time of approximately 4 minutes.

The number of degrees of freedom at the moment of neutron freeze-out is equal to:

$$g_*^{T_n} = 2 + \frac{7}{8} \cdot 4 + \frac{7}{8} \cdot 2 \cdot N_\nu$$

The first contribution arises due to photons, the second due to electrons and positrons, the third is associated with thermalized light neutrinos.

Let's carry out a parallel analysis for the model with three and four neutrinos. Accordingly, for $N_\nu = 3$, $g_*^{T_n} = 10.75$ and for $N_\nu = 4$, $g_*^{T_n} = 12.5$. Although the number of degrees of freedom increases by 16.3%, the rate of the Universe expansion increases by 7.8%, because root dependency.

Indeed, for the radiation domination epoch, the Friedmann equation has the form: $H^2 = \frac{8\pi^3}{90} G g_* T^4$, where the Universe expansion rate is determined by the Hubble parameter, and the plasma cooling rate is determined by the number of degrees of freedom. Thus, the rate of plasma expansion during the neutron freeze-out period increases by 7.8% due to the introduction of an additional light neutrino.

The number of degrees of freedom at the time of nucleosynthesis is

$$g_*^{T_n} = 2 + \frac{7}{8} \cdot 2 \cdot N_\nu \cdot \left(\frac{4}{13}\right)^{4/3}$$

because only two types of relativistic particles affect the plasma expansion rate: photons and neutrinos, and, since neutrinos no longer interact with the plasma matter, their effective contribution is suppressed [11], while electrons and positrons become nonrelativistic and do not contribute to the number of degrees of freedom. Accordingly, for $N_\nu = 3$, $g_*^{T_n} = 3.36$, and for $N_\nu = 4$, $g_*^{T_n} = 3.81$. Although the number of degrees of freedom increases by 13.5%, the Universe expansion rate increases by 6.5%, because root dependency. Thus, the rate of plasma expansion during nucleosynthesis increases by 6.5% when passing from the analysis of the model with three neutrinos to the model with four neutrinos. The average value of the increase factor for the interval from 1.2 s to 265 s is approximately 7%.

Finally, already at the recombination epoch (T = 0.26 eV) and later, a light sterile neutrino with a mass of 2.7 eV becomes nonrelativistic and does not contribute to the number of degrees of freedom. Therefore, the number of degrees of freedom at the moment of recombination is

$$g_*^{rec} = 3.36$$

and remains the same for all following stages of the Universe evolution. This is a very important point that we have already discussed before, but it is useful to emphasize it again. In the meantime, let us calculate the mass content of He$^4$ in two models, i.e. with $N_\nu = 3$ and $N_\nu = 4$.

Using the above well-known calculation scheme, it can be shown that when going from $N_\nu = 3$ to $N_\nu = 4$, the mass fraction of $^4$He increases by 4.9%. The accuracy of calculating the $^4$He mass fraction is much higher if we use the value of the baryon asymmetry and the neutron lifetime. The calculation results are illustrated in Fig. 7, built on the basis of data from [23].

In the Planck 2018 [22], the experimental value of the mass content of $^4$He is fixed at 0.24±0.004 with a measurement accuracy of 2%. However, a wider set of experimental data by year should be presented (Fig. 8, data taken from review [24]), from which it follows that the situation with the experimental value of the $^4$He mass fraction is very uncertain. It is quite obvious that there is a dependence in the observations over the years, which indicates the presence of a systematic effects that have not yet been eliminated, associated with the problem of metallicity (content of heavy elements).

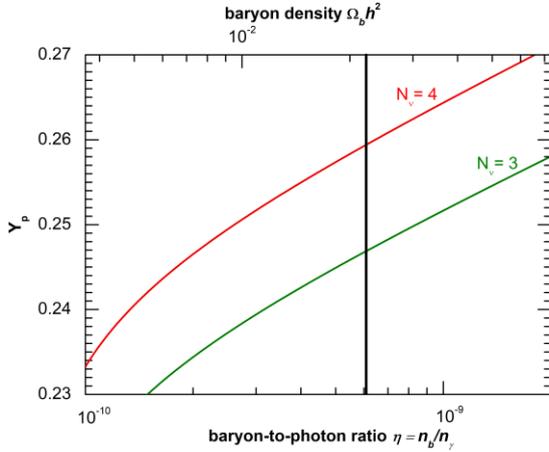

Fig. 7. Y$_p$ abundances as a function of baryon asymmetry at $N_\nu = 3$ and 4 respectively. The line thickness is determined by the experimental accuracy of measuring the neutron lifetime ($\tau_n = 879.4 \pm 0.6$ s). The vertical line corresponds to the value of the baryon asymmetry $(6.090 \pm 0.060) \cdot 10^{-1}$, and its thickness corresponds to one standard deviation. Data taken from [23].

For further analysis, we restrict ourselves to the results of the last two measurements: Izotov 2014 ($Y_P = 0.2551 \pm 0.0022$) [25] and Aver 2015 ($Y_P = 0.2449 \pm 0.0040$) [26]. These results are presented in fig. 9 together with the calculated predictions of the $^4$He mass fraction from the baryon asymmetry and the neutron lifetime.

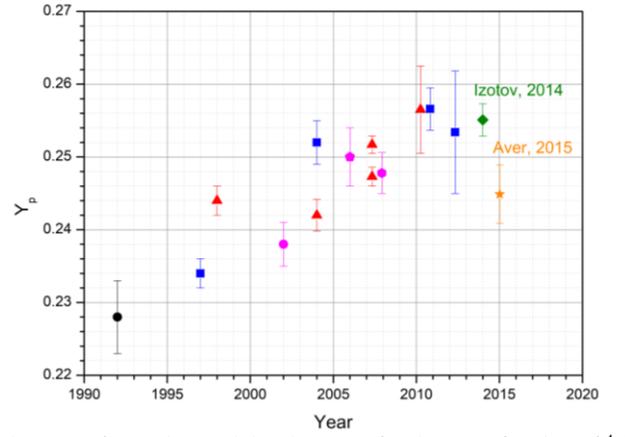

Fig. 8. A set of experimental data by years for the mass fraction of $^4$He. The results of recent measurements are highlighted: Izotov 2014 [25] and Aver 2015 [26]. The rest of the data are taken from [24].

A comparison of the calculated predictions of the mass fraction of $^4$He in the model $N_\nu = 3$ and 4 with the results of experimental measurements is shown in Fig. 9. It can be seen that the measurement results of Aver 2015 are in good agreement with the $N_\nu = 3$, model, and the results of Izotov 2014 are closer to the prediction of the $N_\nu = 4$ model. If we want to compare predictions and astrophysical observations in the $N_\nu = 4$ model, then the discrepancy is 1.9σ according to Izotov and 3.6σ according to Aver.

The analysis of the measurement results for the number of effective degrees of freedom until 2014 looked like it is shown in Fig. 10 obtained from the data from [24] and indicated that $N_{eff} > 3$.

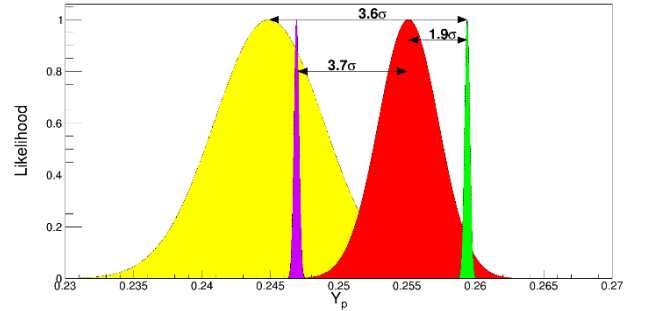

Fig. 9. Comparison of the calculated predictions of the 4He mass abundance, knowing the neutron lifetime and the value of the baryon asymmetry in the model $N_\nu = 3$ and 4 (violet and green peaks, respectively) with the results of astrophysical observations: Izotov 2014 and Aver 2015 (red and yellow distributions, respectively).

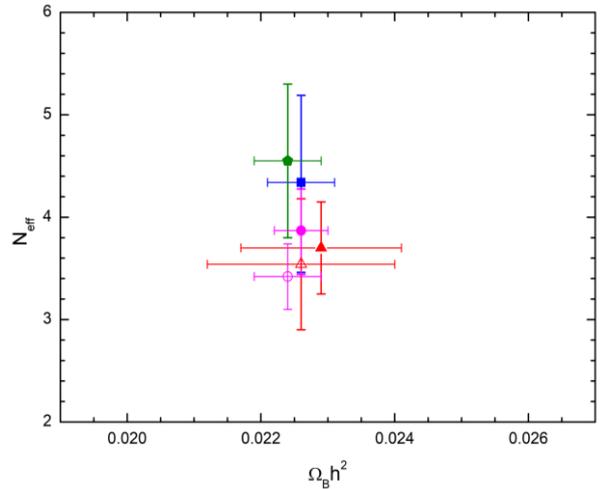

Fig.10 Analysis of measurement results for the number of effective degrees of freedom up to 2014 from [24].

In the publication of the Planck collaboration in 2018, the data of Aver (2015) are used for the mass content of $^4$He, motivating this with a conservative approach.

$$Y_P^{BBN} = 0.247^{+0.017}_{-0.018}, \qquad N_{eff} = 2.89^{+0.31}_{-0.28}$$

The conservatism of Aver's choice of data, apparently, was justified by the fact that in the Standard Model $N_\nu = 3$ and so far, there were no sufficient grounds to discuss the existence of the fourth neutrino.

Below in Fig. 11 we present the results of Aver from [26] and Izotov [25] to give the reader the opportunity to form their own idea of the relationship between the accuracy of the two experiments. Our comment are as follows: 1) individual measurements in the work of Izotov have more than double the accuracy and the range of extrapolation is 1.6 times greater. 2) in Aver's work, the extrapolation result is determined to a large extent by the first and last point, but this is not enough for reliable extrapolation. In our opinion, Izotov's result should be preferred.

Thus, based on the presented astrophysical data, it is impossible to draw a definite conclusion in favor of the model of three or four neutrinos. In any case, one cannot conclude that cosmology, based on astrophysical data, forbids sterile neutrinos with parameters $\Delta m_{14}^2 = 7.3$ eV$^2$ and $\sin^2 2\theta_{14} = 0.36$. Note that the analysis will be continued due to the appearance of new data from EMPRESS collaboration ($Y_P = 0.2370^{+0.0034}_{-0.0033}$ [27]) data on the $^4$He mass fraction.

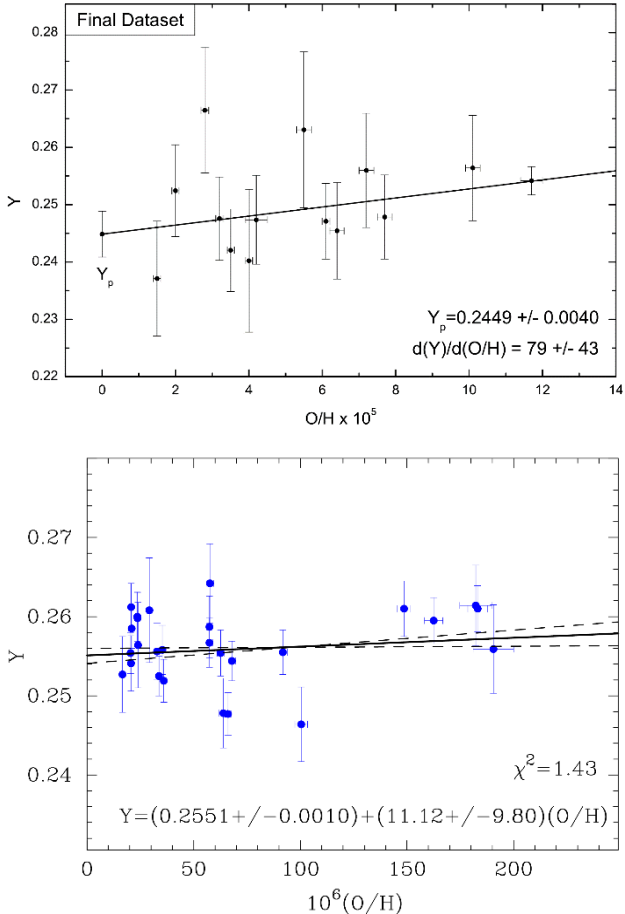

Fig. 11 The results of measuring the $^4$He mass fraction with extrapolation to zero metal content by Izotov 2014 ($Y_P = 0.2551 \pm 0.0022$) and by Aver 2015 ($Y_P = 0.2449 \pm 0.0040$).

**4.4 Hubble constant and number of thermalized neutrinos**

Now let us turn to the analysis of the fact that the Hubble constant and the number of degrees of freedom during the period of neutron freeze-out and nucleosynthesis are related by a simple relation $H_0/\sqrt{g_*}$. Indeed, the plasma temperature is determined by the Universe expansion rate and the plasma cooling rate depending on the number of relativistic degrees of freedom. The fact is that the physical process that occurs in the plasma is determined by its temperature, so if the number of degrees of freedom is increased, then the same temperature will be reached at an earlier stage due to the increased Hubble constant.

One can be considered that an accurate measurement of the Hubble constant from astrophysical data will make it possible to find out how many degrees of freedom there were at that time and make a choice about the number of light thermalized neutrinos.

Unfortunately, there are known contradictions in the question of determining the Hubble constant (Hubble tension). It is discussed that the rate of expansion of the Universe, determined by the Hubble constant $H_0$ and measured by different methods, turned out to be different in magnitude, and the difference reaches 4.4σ. One value $H_0 = 73.2 \pm 1.3$ (km/s)/Mpc (Riess group [28]) refers to the late Universe, and the second value $67.4 \pm 0.5$ (km/s)/Mpc (Planck collaboration [22]) refers to the early Universe. There are also measurements [29] that do not agree with any other results, as shown in Fig. 13, the data for which are taken from [29].

Now let's try to analyze the situation with the Hubble tension within the framework of the cosmological model. In the standard cosmological model, the universe is flat, so you can try to make an analysis considering the curvature of space $\Omega_{Cur}$.

The process of expansion of the Universe depends on time and is described in the cosmological model by the formula:

$$H(z)^2 = H_0^2 \times (\Omega_R \times (1+z)^4 + \Omega_m \times (1+z)^3 + \Omega_\Lambda + \Omega_{Cur} \times (1+z)^2),$$

where $H_0$ - Hubble parameter at the present, $\Omega_R$, $\Omega_m$, $\Omega_0$, $\Omega_\Lambda$ are, respectively, the relative densities of radiation, the density of matter (visible + dark), and the density of dark energy. In addition, there is a term $\Omega_{Cur}$ responsible for the curvature of space, z is the redshift.

Of course, $H_0$ is a constant, but it is extracted by different methods for different values of z and, as a rule, assuming that the term $\Omega_{Cur}$ is equal to zero, i.e. The universe is flat. The formula can be converted to:

$$H^2(z) = H_0^2[\Omega_R(1+z)^4 + \Omega_m(1+z)^3 + \Omega_\Lambda + \Omega_{Cur}(1+z)^2]$$

$$= H_0^2\left[1 + \frac{\Omega_{Cur}(1+z)^2}{\Omega_R(1+z)^4 + \Omega_m(1+z)^3 + \Omega_\Lambda}\right][\Omega_R(1+z)^4 + \Omega_m(1+z)^3 + \Omega_\Lambda]$$

$$H_0^*(z) = H_0\left(1 + \frac{\Omega_{Cur}(1+z)^2}{\Omega_R(1+z)^4 + \Omega_m(1+z)^3 + \Omega_\Lambda}\right)^{1/2}$$

where $\Omega_{Cur}$ is included in the correction factor for $H_0$ if we analyze assuming the Universe is flat, but in fact there is a non-zero curvature.

If we apply such an analysis to the observed Hubble tension: $H_0 = 73.2 \pm 1.3$ (km/s)/Mpc for the late Universe (z<1), and the second value is $67.4 \pm 0.5$ (km/s)/Mpc for the early Universe ( z ~$10^3$ recombination epoch), then we get $\Omega_{Cur} = 0.17$. Figure 12 illustrates the dependence of the effective value $H_0^*$ on z in the presence of space curvature.

$H_0^*(z = 0) = H_0(1 + \Omega_{Cur})^{1/2} = 74, \quad H_0^*(z = 1000) = H_0 = 63$

Such a value of $\Omega_{Cur} = 0.17$ s in obvious contradiction with the limitations of the Planck collaboration ($\Omega_{Cur} = 0.0007 \pm 0.0019$), based on the isotropic distribution of baryon-acoustic oscillations (BAO).

This analysis excludes the explanation of the problem by the curvature of space.

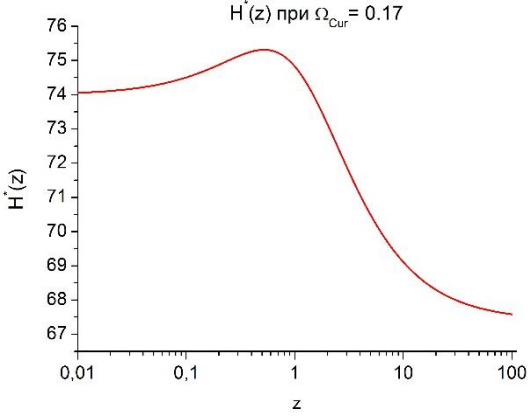

Fig. 12. The dependence of the effective value of $H_0^*$ on z in the presence of space curvature with the value $\Omega_{Cur} = 0.17$.

Continuing the analysis of the data with measurements of the Hubble constant, it is necessary to pay attention to another feature in the measurement of the Hubble constant in the X-ray range [30]. In these studies, it turns out that the Universe is expanding unequally in different directions, i.e. anisotropy is observed in the measurements of the Hubble constant in directions. It is noteworthy that the spread of the values of the Hubble constant is in the same range of 74 - 64 (km / s) / Mpc and is 9%, as in the case of Hubble tension. The choice of coordinate system is determined by the plane of rotation of our galaxy, so we need to take into account the movement of the solar system in it and the movement of our galaxy in the Universe. It is argued in [30] that this is insufficient to explain the observed effect. In the last review article [31], devoted to this issue, a few possible anomalies are considered.

The Hubble tension problem is still open, so we would like to understand the correct value. Let's start again with the fact that the measurement of the Hubble constant by the Riess group refers to the late Universe at $z < 1$. Essentially, these are direct measurements of the Hubble constant. The measurements of the Hubble constant by the Planck collaboration use microwave background data and BBN data. They refer to the early Universe at $z \gg 1$, so the recovery of the true value of $H_0$ is model dependent. It is here that one should look for the causes of contradictions.

Let's go back to the argument that the Hubble constant and the number of degrees of freedom during the period of neutron freeze out and nucleosynthesis are related by a simple relation $H_0/\sqrt{g_*}$, because the plasma temperature is determined by the expansion rate of the Universe and the plasma cooling rate, which depends on the number of relativistic degrees of freedom. It is important to clarify again that the relationship between these quantities for the epoch of radiative dominance is given by the Friedmann equation: $H^2 = \frac{8\pi^3}{90} G g_* T^4$ and hence the strict relation $H_0/\sqrt{g_*}$ arises.

As we already considered in Section 4.2, when passing from the model with three neutrinos to the model with four neutrinos, the number of degrees of freedom during the period of neutron freeze out or neutrino decoupling (1 second) increases from 10.75 to 12.5, i.e. the number of degrees of freedom increases by 16.3%, but the plasma expansion rate increases by 7.8%, since root dependency.

During the period of nucleosynthesis (4 minutes), the degrees of freedom increase from 3.36 to 3.81, i.e. the number of degrees of freedom increases by 13.5%, but the plasma expansion rate increases by 6.5%. Thus, the average value of the increase in the number of degrees of freedom over the interval from 1.2 s to 265 is approximately 7%.

The value of the Hubble constant $H_0 = 67.4 \pm 0.5$ (km/s)/Mpc presented by the Planck collaboration refers to the early Universe and is based on $N_{eff} = 2.96 \pm 0.15$. If we assume that the processes of neutron hardening and nucleosynthesis occurred in the presence of the fourth neutrino, then it is necessary to introduce a correction of 7% for the value $H_0 = 67.4 \pm 0.5$ (km/s)/Mpc.

Then, $H_0^* = 72.1 \pm 0.9$(km/s)/Mpc. Thus, the value from the Planck collaboration ($H_0 = 67.4 \pm 0.5$ (km/s)/Mpc) corrected for the fourth neutrino is $H_0^* = 72.1 \pm 0.9$ (km/s)/Mpcand it is consistent within one standard deviation with the result of the Riess group ($H_0 = 73.2 \pm 1.3$ (km/s)/Mpc).

This is a very important conclusion. The introduction of the fourth neutrino removes the Hubble tension problem. The importance of the conclusion is not just that the Hubble tension problem is removed, but that this required a fourth neutrino..
The thing is that the Planck collaboration deliberately relies on the model of three neutrinos and measurements of the mass content of $^4$He Aver, neglects the results of other measurements.

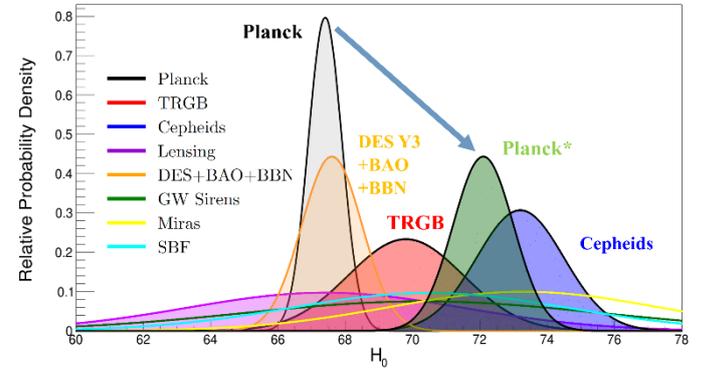

Fig. 13 The introduction of a correction for the fourth neutrino makes it possible to reconcile the value $H_0 = 67.4 \pm 0.5$ (km/s)/Mpc from the Planck collaboration with the value $H_0 = 73.2 \pm 1.3$(km/s)/Mpc from the Riess group. Data taken from [29].

At this point, it is necessary to dwell on the work [32] appeared at the end of last year and posted in the arXiv on February 10th. This paper considers the influence of a sterile neutrino with a mass of 2.7 eV on the Hubble constant using CMB data. Undoubtedly, this is the most important test, without which the analysis of the possibility of introducing sterile neutrinos as dark matter will not be verified.

However, much depends on how the sterile neutrino is found in the composition of dark matter. In our model, a sterile neutrino with parameters $\Delta m_{14}^2 = 7.3$ eV$^2$, $\sin^2 2\theta_{14} = 0.36$ is introduced into the composition of dark matter together with two more heavy sterile ones, which makes it possible to explain the structure of the Universe and bring the contribution of sterile neutrinos to dark matter of the Universe to the level of 27%. It is probably useful to explain here once again that we propose an extension of the Standard Model with right-handed neutrinos - they are sterile. A sterile neutrino with parameters $\Delta m_{14}^2 = 7.3$ eV$^2$, $\sin^2 2\theta_{14} = 0.36$ becomes non-relativistic and is gravitationally drawn into the composition of dark matter even before the recombination epoch. But it manages to participate in the expansion of the plasma at the stage of neutron freeze-out.

The work [32] considers an attempt to fit a sterile neutrino with the parameters $\Delta m_{14}^2 = 7.3$ eV$^2$, $\sin^2 2\theta_{14} = 0.36$ into the existing analysis scheme of the Planck collaboration, varying the parameters to obtain better agreement. The result was that the Hubble tension problem only worsened.

Below we present Table 1 from [32], which shows how, as a sterile neutrino with a mass of 2.7 eV is introduced, the parameters of the Planck collaboration change to obtain the best fitting of the multipole expansion curve for microwave cosmic radiation.

Table 1 from [32]

| Parameter | Planck2018 | $m_s = 2.7\text{eV}$ | | |
|---|---|---|---|---|
| | | $N_{\text{eff}}^s = 0.1$ | $N_{\text{eff}}^s = 0.5$ | $N_{\text{eff}}^s = 1$ |
| $\Omega_{\text{cdm}}$ | $26.45 \pm 0.50$ | $26.79 \pm 0.51$ | $29.35 \pm 0.56$ | $32.36 \pm 0.57$ |
| $\Omega_b$ | $4.93 \pm 0.09$ | $5.07 \pm 0.09$ | $5.49 \pm 0.10$ | $5.88 \pm 0.11$ |
| $\Omega_\nu$ | 0.14 | 1.30 | 4.31 | 7.58 |
| $\Omega_m$ | $31.53 \pm 0.73$ | $33.18 \pm 0.77$ | $39.18 \pm 0.91$ | $45.8 \pm 1.1$ |
| $\Omega_\Lambda$ | $68.47 \pm 0.73$ | $66.82 \pm 0.77$ | $60.82 \pm 0.91$ | $54.2 \pm 1.1$ |
| $H_0$ | $67.36 \pm 0.54$ | $66.50 \pm 0.53$ | $64.07 \pm 0.51$ | $62.20 \pm 0.53$ |

The fact is that in our model there is no need to do anything of the sort. The entire amount of dark matter is the same as before, but due to the fact that a light sterile neutrino thermalizes and changes the number of degrees of freedom. This must be taken into account, as we suggested earlier. As a result, the Hubble constant is corrected in accordance with the Friedmann equation: $H^2 = \frac{8\pi^3}{90} G g_* T^4$, to maintain the neutron freeze-out temperature. The relation $H_0/\sqrt{g_*}$ is an invariant.

Heavy sterile neutrinos (for example, with a mass greater than 50 keV) are not thermalized (see Fig. 4) and do not contribute to the number of degrees of freedom. A light sterile neutrino that thermalizes affects the expansion rate, but does not contribute to nucleosynthesis reactions, because does not interact directly.

The ratio of cosmological parameters for the scheme of dark matter from sterile neutrinos is presented in the last column of Table 2. It is taken the same as in the Planck collaboration, except for the number of degrees of freedom for the period of neutron freeze-out.

Table 2 for the scheme of dark matter with one light and two heavy sterile neutrinos.

| Parameter | Planck2018 | $N_\nu = 3$ $g_*^3 = 10.75$ | $N_\nu = 4$ $g_*^3 = 12.5$ |
|---|---|---|---|
| $\Omega_{\text{cdm}}$ | $26.45 \pm 0.50$ | $26.45 \pm 0.50$ | $26.45 \pm 0.50$ |
| $\Omega_b$ | $4.93 \pm 0.09$ | $4.93 \pm 0.09$ | $4.93 \pm 0.09$ |
| $\Omega_\nu$ | 0.14 | 0.14 | 0.14 |
| $\Omega_m$ | $31.53 \pm 0.73$ | $31.53 \pm 0.73$ | $31.53 \pm 0.73$ |
| $\Omega_\Lambda$ | $68.47 \pm 0.73$ | $68.47 \pm 0.73$ | $68.47 \pm 0.73$ |
| $H_0$ | $67.36 \pm 0.54$ | $67.36 \pm 0.54$ | $\mathbf{72.1 \pm 0.9}$ |

Thus, the temperature power spectrum is still successfully described.

So, the introduction of the fourth neutrino removes the Hubble tension problem without causing contradictions with the CMB observed by the Planck collaboration.

**4.5 Role of lepton asymmetry in BBN analysis**

The earlier consideration of primordial nucleosynthesis was carried out under the assumption that the lepton asymmetry is as small as the baryon asymmetry. However, this condition may be violated.

At the beginning of BBN, neutrons and protons are in equilibrium until the equilibrium is disturbed by a weak interaction. If the $p + \bar{\nu}_e \to n + e^+$ process is suppressed with respect to the $n + \nu_e \to p + e^-$ process due to a smaller number of electron antineutrinos, then this suppresses the neutron-proton ratio and, as a result, $Y_P$ decreases. This decrease in $Y_P$ can be compensated by increasing the number of degrees of freedom $N_\nu^{\text{eff}}$ to keep the same value of $Y_P$. Thus, the presence of lepton asymmetry disguises the presence of the fourth neutrino.

In our calculations in Section 2, it was assumed that the value of lepton asymmetry is negligibly small $10^{-9}$, and the values of chemical potentials can be neglected. If these conditions are not satisfied, then the first-order contribution in the constant $G_f$ must be taken into account in the neutrino potential. The first-order contribution has the form [12]: $V_f = 0,95 \times G_f \eta T^3$, where $\eta$ – is the value of lepton asymmetry. This additional contribution does not depend on the neutrino energy and depends on the temperature as $T^3$. Together with a small value of asymmetry, at high temperatures the contribution of the second order in $G_f$ turns out to be dominant, therefore, in calculations for standard cosmology, the contribution of the first order can be neglected.

If we consider a sufficiently large asymmetry, then the diabatic energy levels of active and sterile neutrinos can intersect, which will lead to resonant oscillations into a sterile state, by analogy with resonant oscillations between electron and muon neutrinos in the Sun (MSW resonance).

Considering the potentials of the form:
$$V_e = 0,95 \times G_f \eta T^3 - 3.5 \times 25 \times G_f^2 \times T^4 \times E$$
$$V_s = 0$$

for various values of the cosmological charge asymmetry - $\eta$ we obtain the dependences of the ratio of the number density of sterile and active neutrinos, which are shown in Fig.14. The value of the cosmological charge asymmetry of the plasma is related to the lepton asymmetry $L_{\nu_\alpha}$ by the factor 4/11 ($\eta_{\nu_\alpha} = 4L_{\nu_\alpha}/11$). The factor is related to the change in the normalization to gamma quanta after the annihilation of electron-positron pairs [33].

At the standard value for cosmology $\eta = 10^{-9}$ the first-order term contribution has no effect on the neutrino density dynamics, which is consistent with the hypothesis that this contribution can be neglected. It turned out that even with an increase in the asymmetry up to $\eta = 10^{-7}$ the contribution of first-order in $G_f$ term does not affect the thermalization dynamics. However, for the values $\eta = 10^{-1}$ and $\eta = 1$ the fraction of sterile neutrinos with respect to electron neutrinos is 0.1 and 0.01, respectively, in the time interval 1 – 100 s (Fig. 14). Therefore, in these cases, sterile neutrinos have little effect on nucleosynthesis, and the contribution of sterile neutrinos with parameters $\Delta m_{14}^2 \approx 7.3 \text{ eV}^2$ and $\sin^2 2\theta_{14} \approx 0.36$ to the dark matter will be 0.5% and 0.05%, respectively. Thus, the presence of a large lepton asymmetry suppresses the presence of the fourth neutrino. However, as will be shown below, $L_e < 0.02$, i.e. $\eta < 0.007$ and therefore there should be no significant screening for neutrinos with experimental parameters.

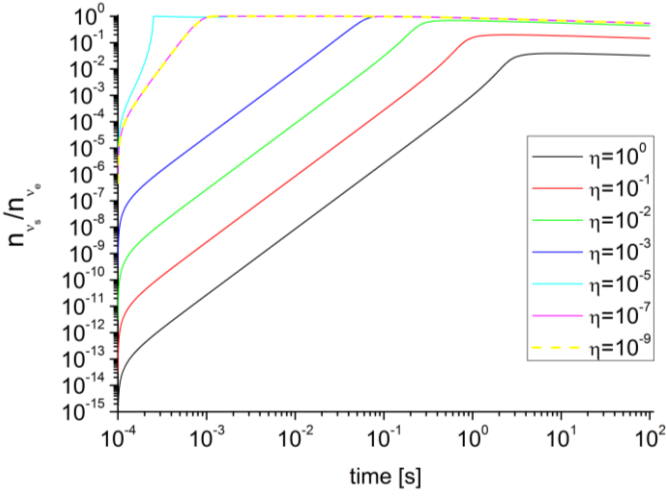

Fig. 14. Dependence of the ratio of the number of sterile neutrinos to the number of electron neutrinos on time for several values of the parameter $\eta$.

The effect of lepton asymmetry on sterile neutrino oscillations was considered by A.D. Dolgov in his 2001 paper [33] "Neutrino oscillations in the early Universe. Resonant case." The idea of this work is that the transformation of $\nu_\alpha$ into $\nu_s$, may be somewhat more favorable than the transformation of the corresponding antineutrinos. The feedback effect is positive and leads to a further increase of the asymmetry and makes the transformation $\nu_\alpha \to \nu_s$ more and more efficient compared to $\bar{\nu}_\alpha \to \bar{\nu}_s$. The lepton asymmetry generated in the early Universe by neutrino oscillations on sterile partners reaches asymptotic values of asymmetry at the level of 0.2 – 0.3 [33]. In papers [12,34], the influence of the chemical potential of neutrinos (antineutrinos) – $\mu_{\nu_a}$ on the magnitude of lepton asymmetry is considered, which affects both primary nucleosynthesis and microwave cosmic radiation. And as a consequence, with non-zero lepton asymmetry, an increase in the effective number of neutrinos - $N_\nu^{\text{eff}}$ is allowed.

In [34] and a number of previous works, the influence of lepton asymmetry on $Y_P$, $N_\nu^{\text{eff}}$ and, ultimately, on the finding of the Hubble constant was also considered.

At a non-zero chemical potential, the Fermi–Dirac distributions for neutrinos (antineutrinos) are written as

$$f_{\nu_a}(p,\xi_a) = \frac{1}{\exp\left(\frac{p}{T_\nu}-\xi_a\right)+1}, \ f_{\bar{\nu}_a}(p,\xi_a) = \frac{1}{\exp\left(\frac{p}{T_\nu}+\xi_a\right)+1},$$

where $\xi_a = \mu_{\nu_a}/T_\nu$,

and $\mu_{\nu_a}$ is the chemical potential, index a corresponds to the active neutrino type $(e,\mu,\tau)$. The value of the lepton asymmetry of the electronic type is given by the equation:

$$L_e \stackrel{\text{def}}{=} \frac{n_{\nu_e}-n_{\bar{\nu}_e}}{n_\gamma} = \frac{1}{36\zeta(3)}\left(\frac{T_{\nu_e}}{T_\gamma}\right)^3\left(\pi^2\xi_{\nu_e}+\xi_{\nu_e}^3\right)$$

If there is a temperature equilibrium $T_{\nu_e} = T_\gamma$ then $L_e = 0.23\ \xi_a$. Using the results of this work, we present in Fig. 15 dependence of $Y_P$ on $\xi$, supplementing the figure with new experimental information about $Y_P$.

Having the exact value of $Y_P$, $N_\nu^{\text{eff}}$ and $H_0$ one can calculate the lepton asymmetry. However, the accuracy of the measurement and the scatter of the $Y_P$. measurements make this impossible. In fact, the introduction of lepton asymmetry into consideration complicates the finding of $N_\nu^{\text{eff}}$. We can only discuss the effect of lepton asymmetry on the attempt to find the number of neutrinos. As can be seen from Fig. 15, as $\xi$ increases, $Y_P$ decreases, thereby masking the presence of the fourth neutrino.

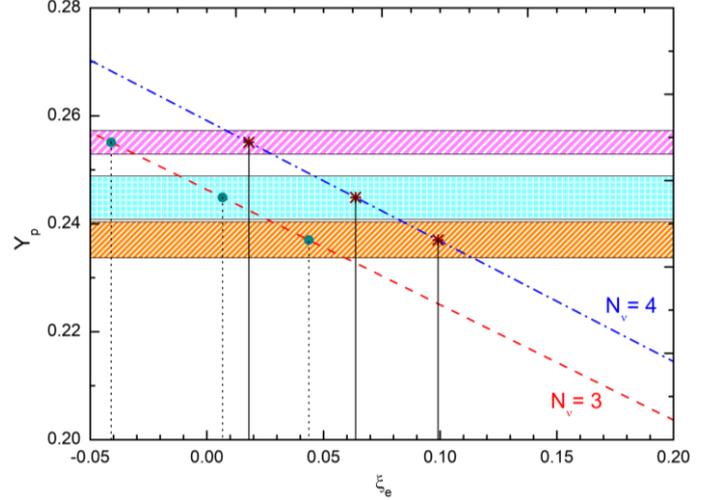

Fig.15. Calculated dependence of He4 mass content $Y_P$ on $\xi$ characterizing lepton asymmetry for $N_\nu = 3$ and 4. $Y_P$ measurements: Izotov 2014 (magenta), Aver 2015 (azure) and latest new $Y_P$ measurements: EMPRESS 2022 [27] (orange). The data for the figure are taken from [34].

For a given set of experimental data on $Y_P$, one can also propose a whole set of estimates for $\xi$ and, accordingly, for the lepton asymmetry $L_e$. Therefore, we can only talk about the upper limit on these values. From Fig.15, one can make constraints on $\xi$, namely $\xi < 0.1$ or on the lepton asymmetry $L_e < 0.02$.

### 4.6 Lepton asymmetry and CP violation

Now let us turn to the consideration of the question of the possible causes of the appearance of lepton asymmetry. Since lepton asymmetry is an inequality in the number of neutrinos and antineutrinos, it is natural to assume that its occurrence can be associated with the violation of CP invariance in neutrino oscillations, moreover, in the inequality of active neutrino oscillations into sterile neutrinos and active antineutrino oscillations into sterile antineutrinos. We propose to consider a scheme when the root cause of the difference between $\nu_\alpha \to \nu_s$ and $\bar{\nu}_\alpha \to \bar{\nu}_s$ is CP violation, and then the resonance amplification mechanism discussed earlier comes into force.

In this regard, it should be noted that in the T2K experiment [35], at the level of two standard deviations, the effect of CP violation in neutrino oscillations is observed. The data obtained indicate the maximum CP violation, the $\delta_{CP}$ parameter is close to -90º. The probability of the transition of muon neutrinos to electron neutrinos exceeds the probability of the transition of muon antineutrinos to electron antineutrinos, which indicates a positive sign of lepton asymmetry. The transition from muon to electron neutrinos can occur through oscillations into a sterile neutrino and further from a sterile neutrino into an electron neutrino. Note that in the neutrino model of four neutrinos, it becomes possible to introduce new CP-violating phases.

If CP violation in the lepton sector occurs through oscillations of active neutrinos into sterile neutrinos, then it is better to understand the purpose of sterile neutrinos in an extended version of the Standard Model, namely, in the production of dark matter and the creation of asymmetry of matter and antimatter. The picture looks suspiciously beautiful! We need a deep theoretical analysis and, of course, an increase in the accuracy of experiments.

## 5. Conclusions

1) A brief analysis of the result of the Neutrino-4 experiment and the results of other experiments on the search for a sterile neutrino is presented. It is noted that a joint analysis of the results of the Neutrino-4 experiment and the data of the GALLEX, SAGE and BEST experiments confirm the parameters of neutrino oscillations declared by the Neutrino-4 experiment (7.3 eV$^2$ and $\sin^2 2\theta_{14} \approx 0.36$ ) and increases the confidence level to 5.8σ.
2) Estimation of the contribution of sterile neutrinos with the mentioned above parameters to the energy density of the Universe is 5%.
3) Extension of the neutrino model by introducing two more heavy sterile neutrinos in accordance with the number of types of active neutrinos will make it possible to explain the structure of the Universe and bring the contribution of sterile neutrinos to the dark matter of the Universe to the level of 27%.
4) It is shown that, based on modern astrophysical data, it is impossible to draw a definite conclusion in favor of the model of three or four neutrinos.
5) It is noted that the introduction of the fourth neutrino eliminates the Hubble tension problem.
6) The influence of lepton asymmetry on the comparison of models of three or four neutrinos is considered.
7) The estimate was made for the upper limit of lepton asymmetry $L_e < 0.02$.
8) The possibility of the appearance of lepton asymmetry due to CP violation during oscillations into sterile neutrinos is discussed.

At the end of the article, we can once again emphasize that the scatter of experimental data in $Y_P$ does not allow us to conclude in favor of the model of three or four neutrinos. Further development of experiments will take place in favor of certain experimental results of measurements. In this regard, the following scenarios can be assumed. Scenario 1. The results of new $Y_P$ measurements give values in favor of Izotov. Then the Planck Collaboration should revise the analysis of the data to finding the Hubble constant, i.e. take into account the fourth neutrino. In this analysis, the results of the Planck collaboration will approximate those of Riess. Scenario 2. New measurements confirm Aver's $Y_P$ measurement, then the Hubble tension problem remains open. Scenario 3. New measurements will confirm the results of $Y_P$ EMPRESS measurements, then it will be necessary to discuss the role of the large lepton asymmetry.

The work was supported by the Russian Science Foundation (Project No. 20-12-00079)
The authors are grateful to V.A. Rubakov and A.D. Dolgov for advice and comments on the theoretical and astrophysical aspects of this work. The authors are grateful to the colleagues of NRC KI PNPI and INR RAS for useful discussions at the seminars.